\documentclass[12]{article}
\usepackage{fleqn}
\usepackage{graphicx}
\begin{document}
\Large
\begin{center}{\bf
The Projective Line Over the Finite Quotient Ring GF(2)[$x$]/$\langle x^{3} - x \rangle$ and
Quantum Entanglement\\
I. Theoretical Background
%Mermin's Pentagram
}
\end{center}
\vspace*{.2cm}
\begin{center}
Metod Saniga$^{\dag}$ and Michel Planat$^{\ddag}$
\end{center}
\vspace*{.1cm} \normalsize
\begin{center}
$^{\dag}$Astronomical Institute, Slovak Academy of Sciences\\
SK-05960 Tatransk\' a Lomnica, Slovak Republic\\
(msaniga@astro.sk)

\vspace*{.1cm}
 and

\vspace*{.1cm} $^{\ddag}$Institut FEMTO-ST, CNRS, D\' epartement LPMO, 32 Avenue de
l'Observatoire\\ F-25044 Besan\c con, France\\
(planat@lpmo.edu)
\end{center}

\vspace*{.0cm} \noindent \hrulefill

\vspace*{.1cm} \noindent {\bf Abstract}

\noindent The paper deals with the projective
line over the finite factor ring $R_{\clubsuit} \equiv$
GF(2)[$x$]/$\langle x^{3} - x \rangle$. The line is endowed with 18 points,
spanning the neighbourhoods of three pairwise distant points. As
$R_{\clubsuit}$ is not a local ring, the neighbour (or parallel)
relation is not an equivalence relation so that the sets of
neighbour points to two distant points overlap. There are nine
neighbour points to any point of the line, forming three disjoint
families under the reduction modulo either of two maximal ideals of the ring. Two of
the families contain four points each and they swap their roles
when switching from one ideal to the other; the points of the one
family merge with (the image of) the point in question, while the
points of the other family go in pairs into the remaining two
points of the associated ordinary projective line of
order two. The single point of the remaining family is sent to the
reference point under both the mappings and its existence stems from
a non-trivial character of the Jacobson radical,
${\cal J}_{\clubsuit}$, of the ring. The factor ring
$\widetilde{R}_{\clubsuit} \equiv R_{\clubsuit}/ {\cal J}_{\clubsuit}$ is
isomorphic to GF(2) $\otimes$ GF(2). The projective line over
$\widetilde{R}_{\clubsuit}$ features nine points, each of them
being surrounded by four neighbour and the same number of
distant points, and any two distant points share two neighbours.
These remarkable ring geometries are surmised to be of
relevance for modelling entangled qubit states, to be discussed in
detail in Part II of the paper.
\\

%\vspace*{.15cm} \noindent {\bf MSC Codes:} 51C05 -- 81R99 -- 81Q99

%\vspace*{.05cm} \noindent {\bf PACS Numbers:} 02.10.Hh -- 02.40.Dr -- 03.65.Ca

%\vspace*{.05cm}
\hspace*{2.cm}Quantum Entanglement

\vspace*{-.1cm} \noindent \hrulefill

\vspace*{.3cm}  \noindent
\section{Introduction}
Geometries over rings instead of fields have been investigated by numerous authors for a long time \cite{tv}, yet they have only recently
been employed in physics  \cite{spl1} and found their potential applications in other natural sciences as well \cite{spl2}. The most prominent, and
at first sight rather counter-intuitive, feature of ring geometries (of dimension two
and higher) is the fact that two distinct points/lines need not have a unique connecting line/meeting point \cite{vk81}--\cite{vk95}.
Perhaps the most elementary, best-known and most thoroughly studied ring geometry is a finite projective Hjelmslev plane \cite{spl1},
\cite{h}--\cite{dj}.

Various ring geometries
differ from each other essentially by the properties imposed on the underlying ring of coordinates. In the present paper we study the structure of
the projective line defined over a finite quotient ring $R_{\clubsuit} \equiv$ GF(2)[$x$]/$\langle x^{3} - x \rangle$. Such a ring is, like those
employed in \cite{spl1} and \cite{spl2}, close enough to a field
to be handled effectively, yet rich enough in its structure of zero-divisors for the corresponding geometry to be endowed with a non-trivial
structure when compared with that of field geometries and to yield interesting and important applications in quantum physics,
dovetailing nicely with those discussed in \cite{spl1} and \cite{spl2}.

\section{Basics of Ring Theory}
In this section we recollect some basic definitions and properties
of rings that will be employed in the sequel and to the extent
that even the reader not well-versed in the ring theory should be
able to follow the paper without the urgent need of consulting
further relevant literature (e.g., \cite{fr}--\cite{ra}).

A {\it ring} is a set $R$ (or, more specifically, ($R,+,*$)) with
two binary operations, usually called addition ($+$) and
multiplication ($*$), such that $R$ is an abelian group under
addition and a semigroup under
multiplication, with multiplication being both left and right
distributive over addition.\footnote{It is customary to denote
multiplication in  a ring simply by juxtaposition, using $ab$ in
place of $a*b$, and we shall follow this convention.} A ring in
which the multiplication is commutative is a commutative ring. A
ring $R$ with a multiplicative identity 1 such that 1$r$ = $r$1 = $r$
for all $r \in R$ is a ring with unity. A ring containing a finite
number of elements is a finite ring. In what follows the word ring
will always mean a commutative ring with unity.

An element $r$ of the ring $R$ is a {\it unit} (or an invertible element) if there exists an element $r^{-1}$ such that $rr^{-1} = r^{-1} r=1$.
This element, uniquely determined by $r$,  is called the multiplicative inverse of $r$. The set of units forms a group under multiplication.
A (non-zero) element $r$ of $R$ is said to be a (non-trivial) {\it zero-divisor} if there exists $s \neq 0$ such that $sr= rs=0$. An element
of a finite ring is either a
unit or a  zero-divisor. A ring in which every non-zero element is a unit is a {\it field}; finite (or Galois) fields, often denoted by GF($q$),
have $q$ elements and exist only for $q = p^{n}$, where $p$ is a prime number and $n$ a positive integer.
The smallest positive integer $s$ such that $s1=0$, where $s1$ stands for $1 + 1 + 1 + \ldots + 1$ ($s$ times), is called the
{\it characteristic} of $R$; if $s1$ is never zero, $R$ is said to be of characteristic zero.

An {\it ideal} ${\cal I}$ of $R$ is a subgroup of $(R,+)$ such
that $a{\cal I} = {\cal I}a \subseteq {\cal I}$ for all $a \in R$.
An ideal of the ring $R$ which is not contained in any other ideal
but $R$ itself is called a {\it maximal}  ideal. If an ideal is of
the form $Ra$ for some element $a$ of $R$ it is called a {\it
principal} ideal, usually denoted by $\langle a \rangle$. A ring
with a unique maximal ideal is a {\it local} ring. Let $R$ be a
ring and ${\cal I}$ one of its ideals. Then $\overline{R} \equiv
R/{\cal I} = \{a + {\cal I} ~|~ a \in R\}$ together with addition
$(a + {\cal I}) + (b + {\cal I}) = a + b +  {\cal I}$ and
multiplication $(a + {\cal I})(b + {\cal I}) = ab +  {\cal I}$ is
a ring, called the quotient, or factor, ring of $R$ with respect to
${\cal I}$; if ${\cal I}$ is maximal, then $\overline{R}$ is a
field. A very important ideal of a ring is that represented by the
intersection of all maximal ideals; this ideal is called the {\it
Jacobson radical}.

A mapping $\pi$:~ $R \mapsto S$ between two rings $(R,+,*)$ and
$(S,\oplus, \otimes)$ is a ring {\it homomorphism} if it meets the
following constraints: $\pi (a + b) = \pi (a) \oplus \pi (b)$,
$\pi (a * b) = \pi (a) \otimes \pi(b)$ and $\pi (1) = 1$  for any
two elements $a$ and $b$ of $R$. From this definition it is
readily discerned  that $\pi(0) = 0$, $\pi(-a) = -\pi(a)$, a unit
of $R$ is sent into a unit of $S$ and the set of elements $\{a \in
R~ |~ \pi(a) = 0\}$, called the {\it kernel} of $\pi$, is an ideal
of $R$. A {\it canonical}, or {\it natural}, map
$\overline{\pi}$:~$R \rightarrow \overline{R} \equiv R/{\cal I}$
defined by $\overline{\pi}(r) = r + {\cal I}$ is clearly a ring
homomorphism with kernel ${\cal I}$. A bijective ring homomorphism
is called a ring {\it iso}morphism; two rings $R$ and $S$ are
called isomorphic, denoted by $R \cong S$, if there exists a ring
isomorphism between them.

Finally, we mention a couple of relevant examples of rings: a polynomial ring, $R[x]$, viz. the set of all polynomials in one
variable $x$ and with coefficients in a ring $R$, and the ring $R_{\otimes}$ that is a (finite) direct product of rings,
$R_{\otimes} \equiv R_{1} \otimes R_{2} \otimes \ldots \otimes R_{n}$, where the component rings need not be the same.

\section{The Ring $R_{\clubsuit}$ and Its Canonical Homomorphisms}
The ring $R_{\clubsuit} \equiv$
GF(2)[$x$]/$\langle x^{3} - x \rangle$ is, like GF(2) itself, of characteristic two and  consists of the following $\#_{{\rm t}}=8$ elements
\begin{equation}
R_{\clubsuit} = \{0, 1, x, x+1, x^{2}, x^{2} +1=(x+1)^{2}, x^{2} + x, x^{2} + x + 1 \}
\end{equation}
which comprise $\#_{{\rm u}}=2$ units,
\begin{equation}
R_{\clubsuit}^{*} = \{1, x^{2} + x + 1 \},
\end{equation}
and $\#_{{\rm z}}= \#_{{\rm t}} - \#_{{\rm u}} = 6$ zero-divisors,
\begin{equation}
R_{\clubsuit} \backslash R_{\clubsuit}^{*}  = \{0, x, x+1, x^{2}, x^{2} + 1, x^{2} + x \}.
\end{equation}
The latter form two principal---and maximal as well---ideals,
\begin{equation}
{\cal I}_{\langle x \rangle} \equiv \langle x \rangle = \{0, x, x^{2}, x^{2} + x\}
\end{equation}
and
\begin{equation}
{\cal I}_{\langle x+1 \rangle} \equiv \langle x + 1 \rangle = \{0, x+1, x^{2}+1, x^{2} + x\}.
\end{equation}
As these two ideals are the only maximal ideals of the ring, its Jacobson radical ${\cal J}_{\clubsuit}$ reads
\begin{equation}
{\cal J}_{\clubsuit} = \langle x \rangle \cap \langle x+1 \rangle = \{0, x^{2} + x \}.
\end{equation}
Recalling that $2 \equiv 0$, and so $+1=-1$, in GF(2), and taking also into account that $x^{3} = x$, the multiplication between the elements
of $R_{\clubsuit}$ is readily found to be subject to the following rules:
\vspace*{0.3cm}
\begin{center}
%\begin{table}
\begin{tabular}{||l|cccccccc||}
\hline \hline
$\otimes$ & 0 & 1 & $x$ & $x^{2}$ & $x+1$ & $x^{2}+1$ & $x^{2}+x$ & $x^{2}+x+1$ \\
\hline
0 & 0 & 0 & 0 & 0 & 0 & 0 & 0 & 0 \\
1 & 0 & 1 & $x$ & $x^{2}$ & $x+1$ & $x^{2}+1$ & $x^{2}+x$ & $x^{2}+x+1$ \\
$x$ & 0 & $x$ & $x^{2}$ & $x$ & $x^{2}+x$ & 0 & $x^{2}+x$ & $x^{2}$ \\
$x^{2}$ & 0 & $x^{2}$ & $x$ & $x^{2}$ & $x^{2}+x$ & 0 & $x^{2}+x$ & $x$ \\
$x+1$ & 0 & $x+1$ & $x^{2}+x$ & $x^{2}+x$ & $x^{2}+1$ & $x^{2}+1$ & 0 & $x+1$ \\
$x^{2}+1$ & 0 & $x^{2}+1$ & 0 & 0 & $x^{2}+1$ & $x^{2}+1$ & 0 & $x^{2}+1$ \\
$x^{2}+x$ & 0 & $x^{2}+x$ & $x^{2}+x$ & $x^{2}+x$ & 0 & 0 & 0 & $x^{2}+x$ \\
$x^{2}+ x+1$ & 0 & $x^{2}+ x+1$ & $x^{2}$ & $x$ & $x+1$ & $x^{2}+1$ & $x^{2}+x$ & 1 \\
\hline \hline
\end{tabular}
%\end{table}
\end{center}

\vspace*{0.4cm}
\noindent
The three ideals give rise to three fundamental quotient rings, all of characteristic two, namely $\widehat{R}_{\clubsuit} \equiv
R_{\clubsuit}/{\cal I}_{\langle x \rangle} = \{0, 1\}$, $\overline{R}_{\clubsuit} \equiv
R_{\clubsuit}/{\cal I}_{\langle x + 1\rangle} = \{0, 1\}$ and
\begin{equation}
\widetilde{R}_{\clubsuit} \equiv
R_{\clubsuit}/ {\cal J}_{\clubsuit} = \{0, 1, x, x+1 \};
\end{equation}
the first two rings are obviously isomorphic to GF(2), whereas the last one
is isomorphic to GF(2)[$x$]/$\langle x^{2} - x \rangle \cong$ GF(2) $\otimes$ GF(2) with componentwise addition
and multiplication (see, e.\,g., \cite{spl2}), as it follows from its multiplication table:
\vspace*{0.3cm}
\begin{center}
%\begin{table}
\begin{tabular}{||l|cccc||}
\hline \hline
$\otimes$ & 0 & 1 & $x$ & $x+1$  \\
\hline
0 & 0 & 0 & 0 & 0  \\
1 & 0 & 1 & $x$ & $x+1$ \\
$x$ & 0 & $x$ & $x$ & 0  \\
$x+1$ & 0 & $x+1$ & 0 & $x+1$  \\
\hline \hline
\end{tabular}
%\end{table}
\end{center}
\vspace*{0.4cm}
\noindent
These quotient rings lead to three canonical homomorphisms $\widehat{\pi}$:~$R_{\clubsuit} \rightarrow \widehat{R}_{\clubsuit}$,
$\overline{\pi}$:~$R_{\clubsuit} \rightarrow \overline{R}_{\clubsuit}$ and
$\widetilde{\pi}$:~$R_{\clubsuit} \rightarrow \widetilde{R}_{\clubsuit}$ of the following explicit forms
\begin{equation}
\widehat{\pi}:~\{0, x, x^{2}, x^{2}+x \} \rightarrow \{0\},~\{1, x+1, x^{2}+1, x^{2}+x+1\} \rightarrow \{1\},
\end{equation}
\begin{equation}
\overline{\pi}:~\{0, x+1, x^{2}+1, x^{2}+x\} \rightarrow \{0\},~\{1, x, x^{2}, x^{2}+x+1\} \rightarrow \{1\},
\end{equation}
and
\begin{eqnarray}
\widetilde{\pi}:&& \{0, x^{2}+x\} \rightarrow \{0\}, \{x, x^{2}\} \rightarrow \{x\}, \{x+1, x^{2}+1\} \rightarrow \{x+1\}, \nonumber \\
&&\{1, x^{2}+x+1\} \rightarrow \{1\},
\end{eqnarray}
respectively.

\section{The Projective Line over $R_{\clubsuit}$ and the Associated Ring-Induced Homomorphisms}
Given a ring $R$ and GL$_{2}$($R$), the general linear group of invertible two-by-two matrices with entries in $R$, a pair ($a, b$) $\in
R^{2}$ is called {\it admissible} over $R$ if there exist $c, d \in R$ such that \cite{her}
\begin{equation}
\left(
\begin{array}{cc}
a & b\\
c & d\\
\end{array}
\right) \in {\rm GL}_{2}(R).
\end{equation}
The projective line over $R$, henceforth referred to as  $PR(1)$, is defined as the set of classes of ordered pairs $(\varrho a, \varrho b)$,
where $\varrho$ is a unit
and $(a, b)$ admissible \cite{her}--\cite{hav}. In the case of $R_{\clubsuit}$, the admissibility condition (10) can be rephrased in simpler terms
as
\begin{equation}
\Delta \equiv \det \left(
\begin{array}{cc}
a & b\\
c & d\\
\end{array}
\right) = ad - bc \in R_{\clubsuit}^{*},
\end{equation}
from where it follows that $PR_{\clubsuit}(1)$ features two algebraically distinct kinds of points: I) the points represented by pairs
where at least one entry is a unit and II) those where both the entries are zero-divisors, not of the same ideal. It is then straightforward to see that
there are altogether
\begin{equation}
\#^{{\rm (I)}} = \frac{\#_{{\rm t}}^{2} - \#_{{\rm z}}^{2}}{\#_{{\rm u}}} = \#_{{\rm t}} + \#_{{\rm z}} = 8 + 6 =14
\end{equation}
points of the former type, namely\\

$~~~(1, 0),~(1, x),~(1, x^{2}),~(1, x+1),~(1, x^{2}+1),~(1, x^{2}+x),~(1, 1),~(1, x^{2}+x+1),$

$~~~(0, 1),~(x, 1),~(x^{2}, 1),~(x+1, 1),~(x^{2}+1, 1),~(x^{2}+x, 1),$\\ \\
and
\begin{equation}
\#^{{\rm (II)}} = \frac{\#_{{\rm z}}^{2} - \#_{{\rm s}}}{\#_{{\rm u}}} = \frac{6^{2} - (2 \times 4^{2} - 2^{2})}{2} = 4
\end{equation}
of the latter type, viz.\\

$~~~(x, x+1) \sim (x^{2}, x+1), ~(x, x^{2}+1) \sim (x^{2}, x^{2}+1),$

$~~~(x+1, x) \sim (x+1, x^{2}),~(x^{2}+1, x) \sim (x^{2}+1, x^{2});$\\\\
here $\#_{{\rm s}}$ denotes the number of distinct pairs of zero-divisors with both entries in the same ideal. Hence,
$PR_{\clubsuit}(1)$ contains $\#^{{\rm (I)}} + \#^{{\rm (II)}} = 14 + 4 = 18$ points in total.

The points of $PR_{\clubsuit}(1)$ are characterized by two crucial relations, neighbour and distant. In particular, two distinct points $X$: $(\varrho a,
\varrho b)$ and $Y$: $(\varrho c, \varrho d)$ are called {\it neighbour} (or,  {\it parallel}) if $\Delta$ is a {\it zero-divisor}, and {\it distant}
otherwise, i.\,e. if $\Delta$ is a {\it unit}. The neighbour relation is reflexive (every point
is obviously neighbour to itself) and symmetric (i.e. if $X$ is neighbour to $Y$ then also $Y$ is neighbour to $X$), but---as we shall see below---not transitive
(i.\,e. $X$  being neighbour to $Y$ and $Y$ being neighbour to $Z$ does not necessarily mean that $X$ is neighbour to $Z$), for $R_{\clubsuit}$ is {\it not}
a local ring (see, e.\,g., \cite{vk95}, \cite{hav}).
Given a point of $PR_{\clubsuit}(1)$, the set of all neighbour points to it will be called its {\it neighbourhood}.\footnote{To avoid any confusion, the reader
must be warned here
that some authors (e.\,g. \cite{bh2}, \cite{hav}) use this term for the set of {\it distant} points instead.} Let us find the cardinality and ``intersection" properties
of this remarkable set. To this end in view, we shall pick up three distinguished pairwise distant points of the line, $U$: $(1, 0)$, $V$: $(0, 1)$ and $W$:
$(1, 1)$, for which we can readily find the neighbourhoods:
\begin{eqnarray}
U: &&U_{1}: (1, x),~~~~~~U_{2}: (1, x^{2}),~~~~~~~U_{3}: (1, x+1),~U_{4}: (1, x^{2}+1),~U_{0}: (1, x^{2}+x), \nonumber \\
&&U_{5}: (x, x+1),~U_{6}: (x, x^{2}+1),~U_{7}: (x+1, x),~U_{8}: (x^{2}+1, x),
\end{eqnarray}
\begin{eqnarray}
V: &&V_{1}: (x, 1),~~~~~~V_{2}: (x^{2}, 1),~~~~~~V_{3}: (x+1, 1),~~V_{4}: (x^{2}+1, 1),~V_{0}: (x^{2}+x, 1), \nonumber \\
&&V_{5}: (x, x+1), ~V_{6}: (x, x^{2}+1),~   V_{7}: (x+1, x), ~  V_{8}: (x^{2}+1, x),
\end{eqnarray}
and
\begin{eqnarray}
W: &&W_{1}: (1, x),~     W_{2}: (1, x^{2}),~        W_{3}: (1, x+1),~   W_{4}: (1, x^{2}+1),~   W_{0}: (1, x^{2}+x+1), \nonumber \\
&&W_{5}: (x, 1),~      W_{6}: (x^{2}, 1),~        W_{7}: (x+1, 1),~   W_{8}: (x^{2}+1, 1).
\end{eqnarray}
 We readily notice that $U_{i} \equiv W_{i}$ for $i = 1, 2, 3$ and 4, $U_{j} \equiv V_{j}$ for $j = 5, 6, 7$ and 8, and
$V_{k} \equiv W_{k+4}$ for $k = 1, 2, 3$ and 4. Now, as the coordinate system on this line can {\it always} be chosen in such a way that the coordinates
of {\it any} three mutually distant points are made identical to those of $U$, $V$ and $W$, from the last three expressions we discern that the
neighbourhood
of any  point of the line features nine distinct points, the neighbourhoods of any two distant points have four points in common (this property thus implying the
already announced non-transitivity of the neighbour relation) and the neighbourhoods of any three mutually distant points have no element in
common---as illustrated in Figure 1.
\begin{figure}[t]
%\vspace*{8.0cm}
\centerline{\includegraphics[width=8truecm,clip=]{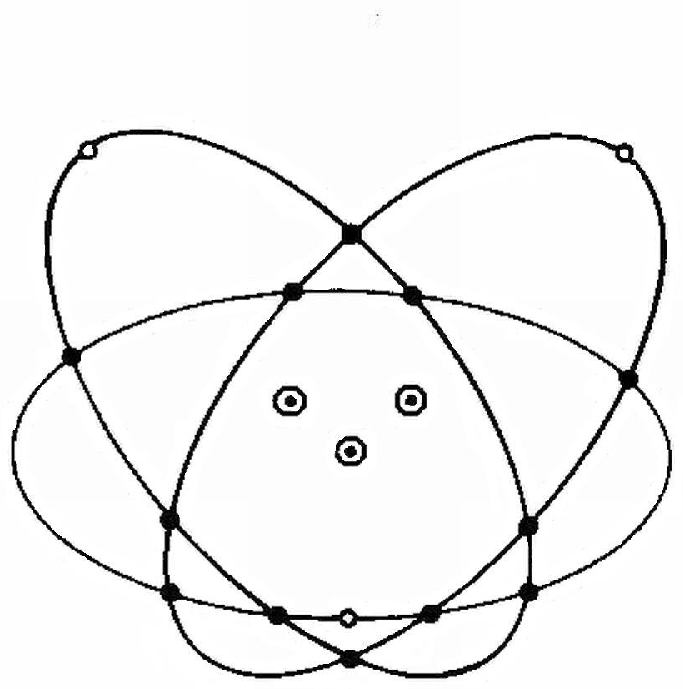}}
\caption{A schematic sketch of the structure of the projective line $PR_{\clubsuit}(1)$. Choosing any three pairwise distant points (represented by
the three double circles), the remaining points of the line are all located on the neighbourhoods of the three points (three sets of points located on
three different ellipses
centered at the points in question). Two neighbourhoods share four points, and as there is no overlapping between the three neighbourhoods,
this way we get twelve points;
the existence of the remaining three points (open circles) is intimately connected with the fact that the ring $R_{\clubsuit}$ has a non-trivial Jacobson
radical.}
\end{figure}

A deeper insight into the structure/properties of neighbourhoods is obtained if we consider the three canonical homomorphisms, Eqs. (8)--(10).
The first two of them induce the homomorphisms from $PR_{\clubsuit}(1)$ into $PG$(1, 2), the ordinary projective line of order two, whilst
the last one induces  $PR_{\clubsuit}(1) \rightarrow P\widetilde{R}_{\clubsuit}(1)$. As $PG$(1, 2) consists of three points, viz.
$U$: $(1, 0)$, $V$: $(0, 1)$ and $W$: $(1, 1)$, we find that the first homomorphism, $PR_{\clubsuit}(1) \rightarrow P\widehat{R}_{\clubsuit}(1)$,
acts on a neighbourhood, taken without any loss of generality to be that of $U$, as follows
\begin{eqnarray}
&&U_{1}, U_{2}, U_{7}, U_{8}, U_{0} \rightarrow \widehat{U}, \nonumber \\
&&U_{5}, U_{6} ~~~~~~~~~~~~~~\rightarrow \widehat{V},\\
&&U_{3}, U_{4} ~~~~~~~~~~~~~~\rightarrow \widehat{W}, \nonumber
\end{eqnarray}
while the second one, $PR_{\clubsuit}(1) \rightarrow P\overline{R}_{\clubsuit}(1)$, shows an almost complementary behaviour,
\begin{eqnarray}
&&U_{3}, U_{4}, U_{5}, U_{6}, U_{0} \rightarrow \overline{U}, \nonumber \\
&&U_{7}, U_{8} ~~~~~~~~~~~~~~\rightarrow \overline{V},\\
&&U_{1}, U_{2} ~~~~~~~~~~~~~~\rightarrow \overline{W}. \nonumber
\end{eqnarray}
The third homomorphism, $PR_{\clubsuit}(1) \rightarrow P\widetilde{R}_{\clubsuit}(1)$, is, however, a more intricate one and in order to fully grasp
its meaning we have first to understand the structure of the line $P\widetilde{R}_{\clubsuit}(1)$.

\begin{figure}[t]
%\vspace*{8.0cm}
\centerline{\includegraphics[width=7.5truecm,clip=]{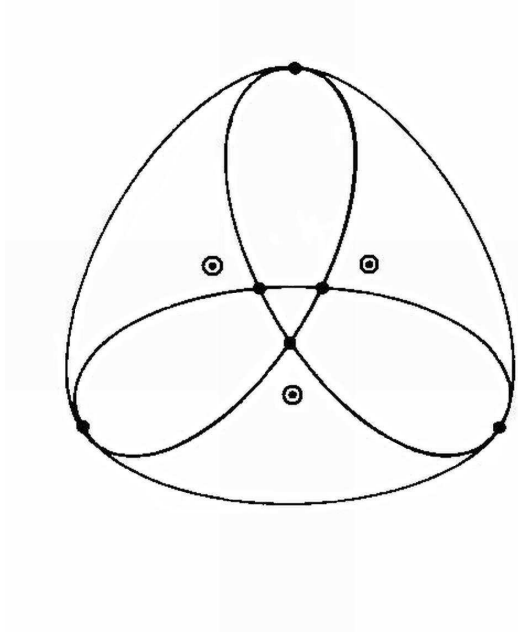}}
\caption{A schematic sketch of the structure of the projective line $P\widetilde{R}_{\clubsuit}(1)$.
As in the previous case, selecting any three pairwise distant points (represented by
the three double circles), the remaining points of the line (filled circles) are all located on the neighbourhoods of the three points
(three sets of points located on three different ellipses centered at the points in question). }
\end{figure}

To this end in view, we shall follow the same chain of reasoning as for $PR_{\clubsuit}(1)$ and with the help of Eq.\,(7) and the subsequent table
find that  $P\widetilde{R}_{\clubsuit}(1)$ is
endowed with nine points, out of which
there are seven of the first kind, \\

$~~~(1, 0),~(1, x),~(1, x+1),~(1, 1),$

$~~~(0, 1),~(x, 1),~(x+1, 1),$\\ \\
and two of the second kind,\\

$~~~(x, x+1),~(x+1, x)$. \\\\
The neighbourhoods of three distinguished pairwise distant points $\widetilde{U}$: $(1, 0)$, $\widetilde{V}$: $(0, 1)$ and
$\widetilde{W}$: $(1, 1)$ here read
\begin{eqnarray}
\widetilde{U}:~~\widetilde{U}_{1}: (1, x),~\widetilde{U}_{2}: (1, x+1),~\widetilde{U}_{3}: (x, x+1),~\widetilde{U}_{4}: (x+1, x),
\end{eqnarray}
\begin{eqnarray}
\widetilde{V}:~~\widetilde{V}_{1}: (x, 1),~\widetilde{V}_{2}: (x+1, 1),~\widetilde{V}_{3}: (x, x+1),~\widetilde{V}_{4}: (x+1, x),
\end{eqnarray}
and
\begin{eqnarray}
\widetilde{W}:~~\widetilde{W}_{1}: (1, x),~ \widetilde{W}_{2}: (1, x+1),~ \widetilde{W}_{3}: (x, 1),~\widetilde{W}_{4}: (x+1, 1).
\end{eqnarray}
From these expressions, and the fact that the coordinates of any three mutually distant points
can again be made identical to those of $\widetilde{U}$, $\widetilde{V}$ and $\widetilde{W}$, we find that the
neighbourhood
of any  point of this line comprises four distinct points, the neighbourhoods of any two distant points have two points in common
(which again implies non-transitivity of the neighbour relation) and the neighbourhoods of any three mutually distant points are disjoint---as
illustrated in Figure 2; note that in this case there exist no ``Jacobson" points, i.\,e. the points belonging solely to a single
neighbourhood, due to the trivial character of the Jacobson radical, ${\cal \widetilde{J}}_{\clubsuit} = \{0\}$. At this point we can already
furnish an explicit expression for
$PR_{\clubsuit}(1) \rightarrow P\widetilde{R}_{\clubsuit}(1)$:
\begin{eqnarray}
&&U_{1}/W_{1}, U_{2}/W_{2} \rightarrow \widetilde{U}_{1}/\widetilde{W}_{1},~~
U_{3}/W_{3}, U_{4}/W_{4} \rightarrow \widetilde{U}_{2}/\widetilde{W}_{2},~~
U_{5}/V_{5}, U_{6}/V_{6} \rightarrow \widetilde{U}_{3}/\widetilde{V}_{3}, \nonumber \\
&&U_{7}/V_{7}, U_{8}/V_{8} \rightarrow \widetilde{U}_{4}/\widetilde{V}_{4},~~
V_{1}/W_{5}, V_{2}/W_{6} \rightarrow \widetilde{V}_{1}/\widetilde{W}_{3},~~
V_{3}/W_{7}, V_{4}/W_{8} \rightarrow \widetilde{V}_{2}/\widetilde{W}_{4},\\
&&U, U_{0} \rightarrow \widetilde{U},~~ V, V_{0} \rightarrow \widetilde{V},~~
W, W_{0} \rightarrow \widetilde{W}. \nonumber
\end{eqnarray}
This mapping will play an especially important role in the physical applications of the theory.

\section{Envisaged Applications of the Two Geometries}
We assume that $P\widetilde{R}_{\clubsuit}(1)$ and
$PR_{\clubsuit}(1)$ provide a suitable algebraic geometrical
setting for a proper understanding of two- and three-qubit states
as embodied in the structure of the so-called Peres-Mermin
``magic" square  and pentagram, respectively \cite{mer}. The
Peres-Mermin square is made of a three-by-three square ``lattice"
of nine 4-dimensional operators (or matrices) with degenerate
eigenvalues $\pm1$.  The three operators in every line/column are
mutually commuting, and each one is the product of the two others
in the same line/column, except for the last column where a minus
sign appears. The algebraic rule for the eigenvalues contradicts
the one for operators, which is the heart of the Kochen-Specker
theorem \cite{ks} for this particular case. The explanation of
this puzzling behaviour is that three lines and two columns have
joint orthogonal bases of {\it un}entangled eigenstates, while the
operators in the third column share a base of {\it maximally
entangled} states. We will establish a one-to-one relation between
the observables in the Peres-Mermin square and the points of the
projective line $P\widetilde{R}_{\clubsuit}(1)$. A closely related
phenomenon occurs in a three-qubit case, with the square replaced
by a pentagram involving ten operators, and the geometrical
explanation will here be based on the properties of the
neighbourhood of a point of the projective line
$PR_{\clubsuit}(1)$. These and some other closely related quantum
mechanical issues will be examined in detail in Part II of the
paper.
\\ \\ \\
\noindent
\Large
{\bf Acknowledgements}
\normalsize

\vspace*{.2cm}
\noindent
The first author thanks Dr.\,Milan Minarovjech for insightful remarks and suggestions,  Mr.\,Pavol Bend\' {\i}k for a careful drawing
of the figures and Dr.\,Richard Kom\v z\' {\i}k for a computer-related
assistance. This work was supported, in part, by the Science and Technology Assistance Agency (Slovak Republic) under the
contract $\#$ APVT--51--012704, the VEGA project $\#$ 2/6070/26 (Slovak Republic) and  the ECO-NET project $\#$ 12651NJ
``Geometries Over Finite Rings and the Properties of Mutually Unbiased Bases" (France).

\end{document}